\documentclass[jcp,twocolumn,showpacs,floats]{revtex4}
\usepackage{graphicx}
\usepackage{amsmath}
\usepackage{tikz}
\usepackage{color}
\graphicspath{{images/},{images/pdf/}}
\begin{document}
\title{Enantioselectivity of (321) chiral noble metal surfaces: A Density Functional Theory study of lactate adsorption}
\author{J.-H.~Franke}
\affiliation{Department of Physics, Campus Plaine - CP 231, Universite Libre de Bruxelles, 1050 Brussels, Belgium} 
\author{D.S.~Kosov}
\affiliation{School of Engineering and Physical Sciences, James Cook University,Townsville, QLD, 4811, Australia}
\date{\today}
\begin{abstract}
The adsorption of the chiral molecule lactate on the intrinsically chiral noble metal surfaces Pt(321), Au(321) and Ag(321) is studied by Density Functional Theory calculations.
We use the oPBE-vdW functional which includes van der Waals forces on an ab initio level. It is shown that the molecule binds via its carboxyl and the hydroxyl oxygen atoms to the surface. The binding energy is larger on Pt(321) and Ag(321) than on Au(321). An analysis of the contributions to the binding energy of the different molecular functional groups reveals that the deprotonated carboxyl group contributes most to the binding energy, with a much smaller contribution of the hydroxyl group. The Pt(321) surface shows considerable enantioselectivity of 0.06 eV. On Au(321) and Ag(321) it is much smaller if not vanishing. The chiral selectivity of the Pt(321) surface can be explained by two factors. First, it derives from  the difference in van der Waals attraction of L- and D-lactate to the surface that we trace to differences in the binding energy of the methyl group. Second, the  multi-point binding pattern for lactate on the Pt(321) surface is sterically more sensitive to surface chirality and also leads to large binding energy contributions of the hydroxyl group. We also calculate the charge transfer to the molecule and the work function to gauge changes in electronic structure of the adsorbed molecule. The work function is lowered by 0.8 eV on Pt(321) with much smaller changes on Au(321) and Ag(321).
\end{abstract}
\maketitle

\section{Introduction}

Molecules that cannot be superimposed on their mirror image are called chiral. This feature is ubiquitous in biologically relevant molecules such as proteins, amino acids and sugars, i.e. the building blocks of life.\cite{Gellman2010} Chiral drug molecules exploit this feature and their chirally specific interactions with these biological molecules leads to vastly different bioactivities.\cite{Sholl2009} To produce enantiopure molecules it is however necessary to either synthesize them in a chirally selective way or to isolate them from a racemic mixture. The first approach requires chirally selective catalysts and the second chirally specific interactions for separation.\cite{Rekoske2001,Blaser2005}

An important approach to enantioseletive catalysis is homogenous catalysis using metal complexes.\cite{Blaser2005} A disadvantage of this approach is that it might involve additional purification processes.\cite{Sholl2009} This could be avoided by using heterogeneous catalysts that require chiral surfaces. These can be obtained by modifying surfaces with chiral molecules or by using intrinsically chiral high Miller index metal surfaces.\cite{Sholl2009,Lorenzo2000,Fasel2006,Kuhnle2002,Mallat2007,Ahmadi1999,Sholl1998,Clegg2011,Kyriakou2011,Meemken2012,Gross2013,Lawton2013} If these surfaces are made from catalytically active metals, they could be expected to endow a well-performing catalyst with chiral selectivity. To achieve this, it is necessary to obtain a fundamental understanding of chirally selective molecule-surface recognition, which requires the use of advanced surface science techniques as well as Density Functional Theory (DFT) calculations.\cite{Eralp2011,Bombis2010,Zhao2004,Greber2006,Horvath2004,Huang2011,Huang2008,Cheong2011,Han2012,Bhatia2005,Bhatia2008,Sljivancanin2002,Schmidt2012,James2008}

Here, we study deprotonated lactic acid - lactate - on intrinsically chiral noble metal surfaces. As a promising molecule for green chemistry\cite{Gallezot2012,Poliakoff2002,Holm2010} lactic acid is already finding industrial use, also in the form of polylactic acid.\cite{Gallezot2012,Ragauskas2006,MadhavanNampoothiri2010,Rasal2010,Katiyar2010} Since the thermochemical properties of polylactic acid depend on the chirality of its monomer constituents enantioselective control is also important for this system.\cite{MadhavanNampoothiri2010,Platel2008}

Specifically we focus here on the adsorption of the two enantiomers of lactate on the catalytically active noble metal surfaces Pt(321), Au(321) and Ag(321). A previous study has shown that the (321) surface enables binding to two consecutive kink atoms which increases chiral selectivity for lactic acid.\cite{Franke2013} Thus, these surfaces can reasonably be expected to maximize chiral selectivity effects also for lactate for each metal. We find that on all surfaces studied the most stable configurations of the molecule indeed optimizes its interaction with kink sites by binding with its three oxygen atoms to the two available kink atoms and the ridge like atom between. This general binding pattern can be achieved by different adsorption geometries, which are very similar in energy. We analyze the different contributions to the binding energy of the carboxylic, hydroxlyic and methyl groups to elucidate trends that might point to a general microscopic mechanism in the adsorption of carboxylic acids with an adjacent hydroxyl group. We find that the Pt(321) surface exhibits a chiral selectivity of 0.06 eV for the adsorption of lactacte, while this effect is smaller or non-existent on Au(321) and Ag(321). The calculated chiral selectivity is large enough to be experimentally detectable and should lead to significant chiral excesses of adsorbed lactate out of a racemic mixture.\cite{Huang2008,Huang2011,Yun2013} 

The remainder of the paper is organized as follows. In Sec. 2, we describe the parameters used in the computations. Section 3 presents the results obtained for adsorption on chiral surfaces, while Section 4 deals with the electronic structure of the different adsorption configurations. Conclusions are given in Section 5.

\section{Computational details}

\begin{figure*}[Htb]
\includegraphics[width=13cm]{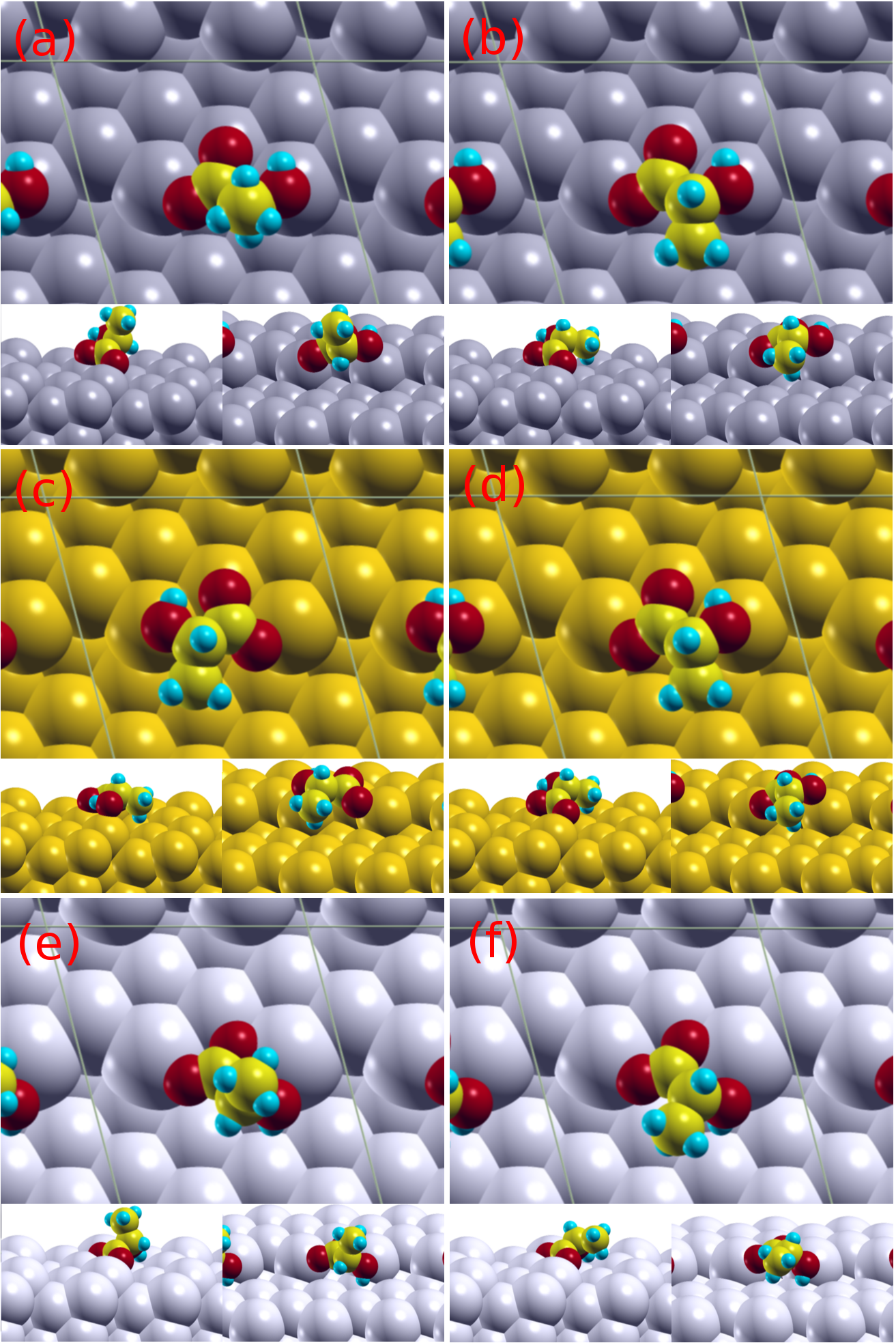}
\caption{L-lactate (\textbf{a},\textbf{c},\textbf{e}) and D-lactate (\textbf{b},\textbf{d},\textbf{f}) adsorbed on Pt(321)$^S$ (\textbf{a},\textbf{b}), Au(321)$^S$ (\textbf{c},\textbf{d}) and Ag(321)$^S$ (\textbf{e},\textbf{f}) surfaces.}
\label{fig:geom111}
\end{figure*}

\begin{table*}[Htb]
\caption{Adsorption energies, work functions and Hirshfeld charges of the two enantiomers of lactate on Pt(321)$^S$, Au(321)$^S$ and Ag(321)$^S$. Numbers is brackets denote the contributions of the non-local correlation energy to the adsorption energy and the change of work function induced by the adsorption of the lactate molecule, respectively.}
\begin{tabular}{c|c|c|c}
\hline
\multicolumn{1}{c|}{lactate on} & \multicolumn{1}{|c|}{adsorption energy oPBE-vdW} & \multicolumn{1}{c}{work function} & \multicolumn{1}{|c}{Hirshfeld charge} \\
 & (eV) & (eV) & of molecule (e)\\
\hline
L on Pt(321)$^S$   & -2.98 (-1.47) & 4.65 (-0.84) & -0.07 \\
D on Pt(321)$^S$   & -3.04 (-1.62) & 4.71 (-0.78) & -0.06 \\
L on Au(321)$^S$   & -2.36 (-1.38) & 4.96 (-0.16) & -0.24 \\
D on Au(321)$^S$   & -2.37 (-1.40) & 5.00 (-0.12) & -0.24 \\
L on Ag(321)$^S$   & -2.98 (-1.18) & 4.28 ( 0.02) & -0.28 \\
D on Ag(321)$^S$   & -2.97 (-1.22) & 4.27 ( 0.01) & -0.28 \\
\hline
\end{tabular}
\label{tab:energ}
\end{table*}

We obtained our results with the DFT code VASP 5.3\cite{Hafner2008,Kresse1996a,Kresse1996b} with the oPBE-vdW functional\cite{Klimes2010,Klimes2011} throughout. The correct description of the van der Waals forces turns out to be crucial since they make up a large part the adsorption energy and, in our particular case, they are ultimately responsible for the appearance of chiral selectivity. We opted for this special version of the vdW-DF functional since we wanted to combine the well-tested PBE class of functionals with an optimal accuracy of the vdW nonlocal correlation. The Projector Augmented Wave method\cite{Bloechl1994,Kresse1999} was employed with valence wave functions expanded up to an energy cutoff of 400 eV. The lattice constants were determined using a K-mesh of 17x17x17 in conjunction with the tetrahedron method with Bloechl corrections. Fitting of a series of fixed volume calculations to the Murnaghan equation of state gave lattice constants of 3.999\AA, 4.180\AA\ and 4.162\AA\ for Pt, Au and Ag, respectively. The force cutoff for all structural relaxations is 10 meV/\AA. For all slab calculations dipole corrections to the potential are applied throughout.\cite{Neugebauer1992} For optimal accuracy all energies given are calculated with evaluation of the projector functions in reciprocal space.

The (321) surfaces were constructed with a thickness corresponding to 6 layers of (111) orientation and the upper half of the slabs were relaxed using 7x7x1 K-meshes, respectively. Lactate was adsorbed on a 2x2 supercell of the (321) surfaces. Due to the increased unit cell size the K-mesh was reduced to 3x3x1 for all surfaces. All molecular degrees of freedom were allowed to relax as was the upper part of the metal slabs. The molecules were adsorbed on the relaxed side of the surfaces which constitutes a (321)$^S$ surface.\cite{Ahmadi1999,Sholl2001} Adsorption energies $E_{adsorption}$ are given with reference to the isolated surface $E_{surface}$ relaxed upon removing the molecule from the unit cell using identical computational parameters and the energy of the molecule $E_{lactate}$

\begin{equation}
E_{adsorption} = E_{mol\ on\ surface} - E_{surface} - E_{lactate}.
\end{equation}

Additionally the same calculations are performed using only the non-local correlation energy of the oPBE-vdW functional, thus yielding an estimate for the contribution of the dispersion forces to the adsorption energy. 

To compare the adsortion energy of lactate, calculated above, with the adsorption energies of lactic acid\cite{Franke2013}  - assuming that the removed H atom forms molecular hydrogen - one needs to add the deprotonation energy $E_{deprot}$ 

\begin{equation}
E_{deprot} = E_{lactate} + \frac{1}{2}E_{H_2} - E_{lactic\ acid}
\end{equation}
to all adsorption energies. This energy is calculated as 2.14 eV. 

\section{lactate on Pt(321), Au(321) and Ag(321)}

\begin{table*}[Htb]
\caption{Adsorption energy component analysis for the chiral surface configurations. Adsorption energies of hydrogen saturated hydroxyl, carboxyl and methyl groups in the frozen adsorption geometries and deformation energies of the substrate and molecule are calculated. The numbers is brackets denote the non-local correlation part in the adsorption energies. The sum of these energy components differs from the overall binding energy (cf. Table \ref{tab:energ}) as a result of the contributions of the neglected carbon and hydrogen atoms and the introduction of additional saturating hydrogen atoms.}
\begin{tabular}{c|c|c|c|c|c|c}
\hline
\multicolumn{1}{c}{lactate on} & \multicolumn{2}{|c}{deformation energy (eV) } & \multicolumn{3}{|c}{adsorption energy (eV)} & \multicolumn{1}{|c}{sum of components (eV)} \\
 & surface & molecule & COH group & COOH group & CH3 group &  \\
\hline
L on Pt(321)$^S$   & 0.13 & 0.32 & -0.45 (-0.45) & -2.90 (-0.98) &  0.02 (-0.10) & -2.88 \\
D on Pt(321)$^S$   & 0.14 & 0.33 & -0.44 (-0.44) & -2.93 (-0.97) & -0.09 (-0.34) & -2.99 \\
L on Au(321)$^S$   & 0.27 & 0.15 & -0.19 (-0.36) & -2.38 (-0.79) & -0.06 (-0.37) & -2.21 \\
D on Au(321)$^S$   & 0.26 & 0.11 & -0.21 (-0.34) & -2.40 (-0.80) & -0.10 (-0.36) & -2.34 \\
L on Ag(321)$^S$   & 0.09 & 0.41 & -0.32 (-0.38) & -3.20 (-0.80) & -0.03 (-0.07) & -3.05 \\
D on Ag(321)$^S$   & 0.08 & 0.45 & -0.33 (-0.38) & -3.20 (-0.78) & -0.09 (-0.15) & -3.09 \\
\hline
\end{tabular}
\label{tab:decomp}
\end{table*}

The general pattern for the most stable adsorption configurations is one of three-fold oxygen binding to two kink atoms and the atom bridging them. For the Pt(321) and Au(321) surfaces, the hydrogen atom of the hydroxyl group is on the upper facet of the surface step. For the most stable Ag(321) surface configurations the hydrogen atom is pointing towards the lower facet. On the Pt(321) and Ag(321) surfaces the position of the oxygen atoms of the lactate molecule is the same for both its chiralities, i.e. the configurations can be converted to one another by exchanging the hydrogen atom and the methyl group on the chiral center of the molecule. On Au(321) the two chiralities of the molecule can be interconverted by vertically mirroring the two configurations.

The binding energies obtained show that the lactate molecule is bound much stronger to the Pt(321) and Ag(321) surfaces than to Au(321). On Pt(321) there is a sizable chiral selectivity of 0.06 eV, while for Au(321) and Ag(321) the calculated chiral selectivity is about vanishing. To understand the role of dispersion forces, we separately calculated the contribution of the non-local correlation energy (i.e. the dispersion or van der Waals interaction energy) to the binding energy. For all surfaces studied this energy contributes about half to the binding energy. It is largest for the Pt(321) surface and smallest for Ag(321). The rest of the binding energy is thus largest for Ag(321) and smallest for Au(321). For D-lactic acid on Pt(321) the van der Waals binding energy is larger by 0.15 eV than for L-lactic acid, indicating that this interaction is behind the calculated chiral selectivity. Accordingly, the non-local correlation binding energy differences between the two chiralities of the molecule on Au(321) and Ag(321) are smaller.

To understand the different binding patterns better and to infer general trends for organic acids with carboxylic acid and hydroxyl groups at their chiral center we carry out a binding energy component analysis.\cite{Sljivancanin2002,Franke2013} To this end we freeze the COO, COH and CH$_3$ groups in their adsorption configuration, remove the rest of the molecule and saturate the groups with additional hydrogen. We calculate the binding energy of the thus obtained groups separately and also calculate the deformation energy of the molecule and substrate by comparing their frozen geometry under removal of the other component with the one obtained in the relaxed configuration prior to adsorption.

We find that on all surfaces the binding energy is strongly dominated by the binding of the COO group. Its binding energy is actually on the order of the binding energy of the whole molecule. More accurately it is about 0.2 eV larger on Ag(321), about the same on Au(321) and 0.1 eV smaller on Pt(321). The binding energy of the COH group is much smaller on all surfaces. It increases from Au(321) through Ag(321) to Pt(321). In addition to the van der Waals interaction, an important reason for the chiral selectivity observed on Pt(321) is probably the relatively similar contributions of the COH binding energy component when compared to the COO one. While on Pt(321) the ratio between the two is about 1:7, it is about 1:10 and 1:12 on Au(321) and Ag(321), respectively. The deformation energies of the molecule and the substrate are larger for the more strongly binding Ag(321) and Pt(321) surfaces than on Au(321). On Ag(321) more energy is lost to deformation of the molecule than on Pt(321) and vice versa for the substrate.  

Most differences in energy components for the two chiralities on a given substrate are small. However, there are some energy contributions that show that the overall chiral selectivity is the sum of many effects. The first are the deformation energies of the molecule and the substrate. On Pt(321) and Ag(321) the deformation energy of the molecule is larger for D-lactate, which has the molecule lying more flat on the surface and being more deformed by the surface interaction. On Au(321) L-lactate is more deformed, which is also the configuration with the more flat-lying molecule. Interestingly, the dispersion contribution to the binding of the COH and COO groups are 0.02 eV larger for L-lactate than for D-lactate on Pt(321) and Ag(321), in contrast to the general trend observed. The impact of the dispersion forces is however much more important in the case of the methyl groups. This can be understood from their different binding geometries on Pt(321) and Ag(321). For L-lactate on Pt(321) and Ag(321) the methyl group is at a larger distance from the surface than for the more flat-lying D-lactate adsorption configuration. Here there are sizeable differences in the dispersion part of the binding energy, that is somewhat mitigated by the local part of the binding energy. On Pt(321) this difference amounts to 0.24 eV in the dispersion part and 0.11 eV overall. On Ag(321) these values are lower at 0.08 eV and 0.06 eV, respectively. For Pt(321) this can explain the large difference in the overall dispersion contribution to the binding energy that in turn leads to the overall observed chiral selectivity. On Ag(321) the dispersion forces are too weak to imply an overall chiral selectivity. For L,D-lactate on Au(321) the positions of the methyl group are similar and thus there appears no strong difference in the dispersion parts of their binding energies. However, a difference of 0.04 eV is present in the overall binding energy. 

\section{Electronic structure}

\begin{figure}[Htb]
\includegraphics[width=8cm]{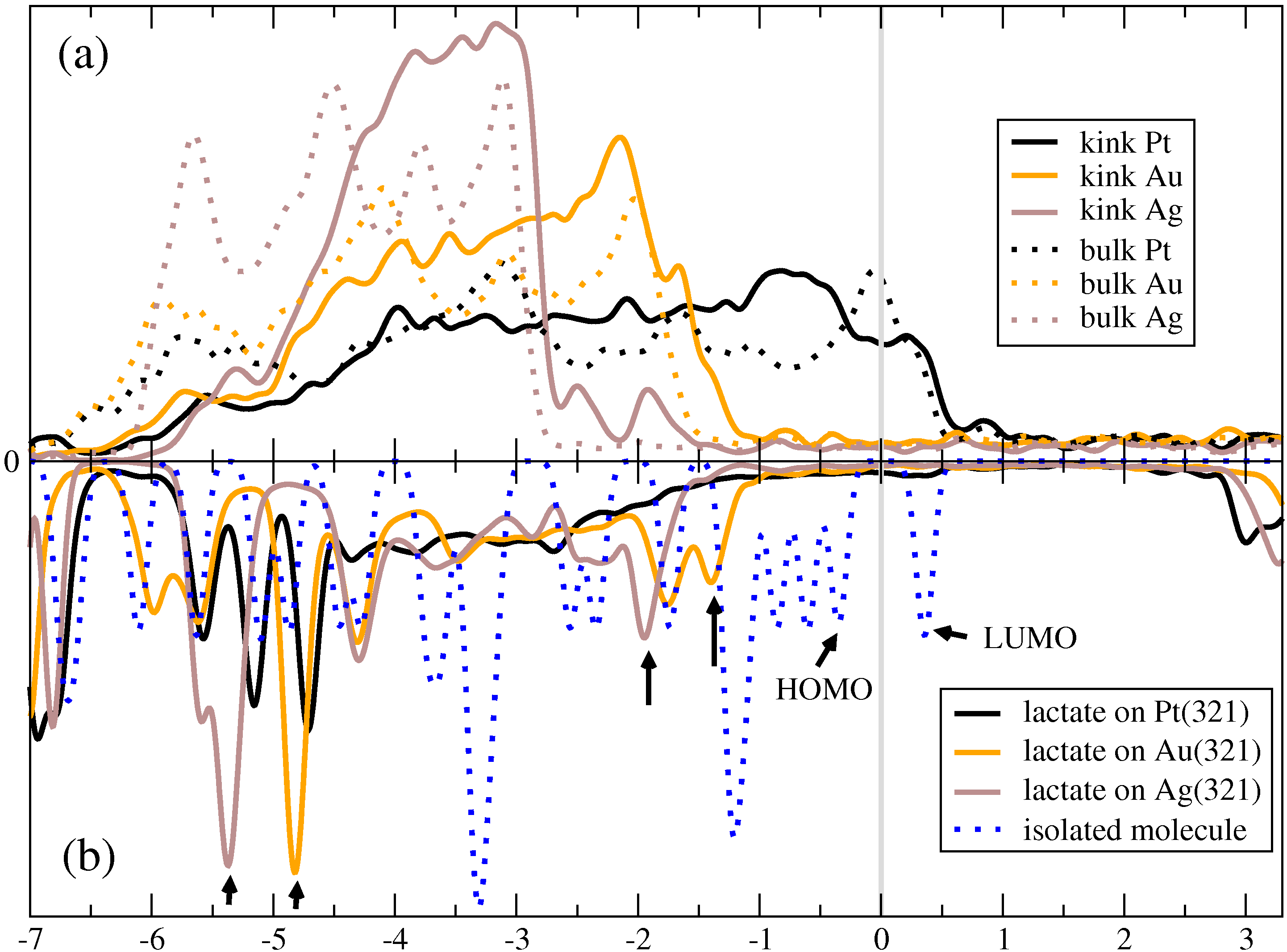}
\caption{Projected Density of states (PDOS) of a bulk and the kink atoms where the carboxyl group is bound to for L-lactate on Pt(321)$^S$, Au(321)$^S$ and Ag(321)$^S$ (\textbf{a}) and PDOS of the corresponding molecules adsorbed on the threee surfaces (\textbf{b}). PDOS of the molecule in vacuum is shown for comparison in blue. PDOS for the D-lactate adsorption configurations is very similar. Features discussed in the text are marked by arrows.}
\label{fig:PDOS}
\end{figure}

We calculated the work functions and the Hirshfeld charges of the two lactate enantiomers for the most stable configurations. It turns out that on Pt(321) the work function is significantly lowered with respect to the value of the pristine substrate while on Ag(321) and Au(321) this modification is much smaller. Also, the absolute Hirshfeld charge (negative in all cases) is much larger on Ag(321) and Au(321) than on Pt(321). It seems that the bigger charge of the molecule on the first two substrates compensates the push-back effect, while the charge is too small to do this on Pt(321).\cite{Bagus2002,Michaelides2003} This points to significant differences in the binding mechanism between Pt(321) on the one side and Au(321) and Ag(321) on the other side. It is interesting to note that the changes in the work function for different chiralities do not reflect the surface enantioselectivity.  For example, the work function differences for D,L-lactate on chirally nonselective Au(321) and chirally selective Pt(321) are comparable at 0.04 eV and 0.06 eV, respectively.

The differences between the substrates are also visible in the Projected Density of States (PDOS) of the adsorbed molecules. Fig. \ref{fig:PDOS}(a) gives the PDOS of the kink atoms the carboxyl group interacts with and compares it to the one of a bulk atom. It is evident that the d-states that make up the bulk of the PDOS between 6 eV binding energy and the Fermi edge are significantly tilted on the kink atoms towards lower binding energies when compared to the bulk atom. This we attribute to the undercoordination of this atom. It is also evident that the low-binding energy cutoff for the d-states moves towards lower binding energies from Pt over Au to Ag. 

The molecular PDOS (cf. Fig \ref{fig:PDOS}(b)) attests to the strong interaction of the molecule with the substrate on all surfaces as the sharp peaks of the molecule in vacuum are significantly broadened. It is also evident that the energetic positioning of the d-states of the kink impacts the energies of the hybridized molecular states for the adsorbed molecule. Thus, going down from the Fermi edge, one finds the onset of molecular PDOS at higher binding energies on Ag(321) than on Au(321). The corresponding PDOS starts at 2 eV binding energy on Ag(321) and 1.3 eV on Au(321). Also a sharper peak is found at about 4.8 eV binding energy on Au(321) and 5.4 eV on Ag(321). These two observations are in agreement with the larger total binding energy on Ag(321). In contrast, on Pt(321) the PDOS of the molecules shows a more gradual onset when going down from the Fermi edge without any sharp peaks at low binding energy. Only between 4.5 and 5.5 eV binding energy are sharp peaks observed. The absence of the peaks at lower energies is also consistent with the smaller Hirshfeld charge of the molecule on Pt(321).

\section{Conclusion}

We studied the adsorption of lactate on Pt(321), Au(321) and Ag(321). On all surfaces the most stable adsorption configurations exhibit interactions of the carboxylic and hydroxylic oxygen atoms with the kink atoms. The molecule is much more strongly bound to Pt(321) and Ag(321) than to Au(321). The carboxyl group strongly dominates the binding energy for all surfaces, with the hydroxyl group providing much less binding energy. Van der Waals forces provide about half of the binding energy on Pt(321), slightly more than that on Au(321) and slightly less on Ag(321).

Pt(321) shows considerable enantioselectivity of 0.06 eV toward the adsorption of lactate. For the Au(321) and Ag(321) surfaces the chiral selectivities are much smaller. Two factors contribute to this outcome: (i) the large difference in dispersion energy between the two enantiomers on Pt(321) that can be traced to the methyl group (ii) the smaller carboxyl/hydroxyl binding energy contribution ratio. These two effects point to a more balanced three-point binding pattern with the constituents carboxyl, hydroxyl and the methyl group on Pt(321). The chiral selectivity on Pt(321) is also much larger for lactate than for lactic acid, which could be used to experimentally distinguish them.\cite{Franke2013} 

The electronic structure of the adsorbed molecules shows that charge transfers to the molecule as well as the work function changes are similar for adsorption on Ag(321) and Au(321) that differ from Pt(321). In the first two cases a charge transfer to the molecule is observed, compensating the push-back effect to yield only small changes to the work function. On Pt(321) the charge transfer is smaller and the work function is lowered by about 0.8 eV. 

The PDOS of the adsorbed molecule attests to the strong hybridization of electronic states between the surfaces and the lactate molecule. An onset of occupied states near the Fermi energy that corresponds to the density of states of the kink atom is observed for lactate on Ag(321) and Au(321). As the d-states are at higher binding energies on the Ag surface, the PDOS of the adsorbed molecule is shifted accordingly, leading to higher binding energies. On Pt(321) the onset of the molecular PDOS is more gradual, despite the availability of electronic states on the kink atom right down from the Fermi edge.

\section{Acknowledgements}

This work has been supported by the Francqui Foundation, and Programme d'Actions de Recherche Concertee de la Communaute Francaise, Belgium. We would like to thank Pierre Gaspard and Thierry Visart de Bocarme for useful discussions. We also acknowledge the Computing Center of ULB/VUB for computer time on the HYDRA cluster.


\end{document}